\documentclass[11pt]{article}

\usepackage[T1]{fontenc}
\usepackage[utf8]{inputenc}
\usepackage{lmodern}
\usepackage{amsmath,amssymb}
\usepackage{graphicx}
\usepackage{booktabs}
\usepackage{array}
\usepackage{caption}
\usepackage{geometry}
\usepackage[round,authoryear]{natbib}
\usepackage{xurl}
\usepackage[table]{xcolor}
\usepackage{placeins}
\usepackage{float}
\usepackage[
  colorlinks=true,
  linkcolor=blue,
  citecolor=blue,
  urlcolor=blue,
  pdfpagemode=UseOutlines,
  bookmarksopen=true,
  bookmarksopenlevel=2
]{hyperref}
\usepackage{microtype}

\graphicspath{{figures/}}
\newcommand{\boldurl}[1]{\href{#1}{\nolinkurl{#1}}}
\newcommand{\bolddoi}[1]{\href{https://doi.org/#1}{\nolinkurl{doi:#1}}}
\definecolor{tablehighlight}{RGB}{204,239,250}
\newcommand{\corrhighlight}[1]{\cellcolor{tablehighlight}#1}
\newcommand{\minititle}[1]{%
  \par\medskip
  \noindent\textbf{#1}%
  \par\noindent
}

\title{\textbf{Web application for retrospective geomagnetic storm detection, main phase analysis and identification of solar wind influence on storm intensity}}
\author{D. Papakonstantinou\textsuperscript{1} \and I. A. Daglis\textsuperscript{1}\\[0.5em]
\small \textsuperscript{1}Department of Physics, National and Kapodistrian University of Athens, Athens, Greece}
\date{}

\begin{document}
\maketitle

\begin{abstract}
\small
Geomagnetic storms are major disturbances of Earth's magnetosphere resulting from variations in the solar wind, leading to significant changes in magnetospheric currents, plasma populations, and electromagnetic fields. This work presents a web application for retrospective detection and analysis of geomagnetic storms using hourly Dst and solar wind data from OMNIWeb. The analysis covers 1964 through May 2026. The application detects storm events, identifies their main phases, computes solar wind derived parameters during the main phase, and examines their relationship with storm intensity. Its detection and main phase identification parameters are user-adjustable, allowing different assumptions to be tested interactively rather than imposed through a single fixed parameter set. Using the default parameter set provided by the application, 1796 storms with $\mathrm{Dst}_{\min} \leq -50$ nT were detected, including 1422 moderate storms ($-100 < \mathrm{Dst}_{\min} \leq -50$ nT), 347 intense storms ($-250 < \mathrm{Dst}_{\min} \leq -100$ nT), and 27 super-storms ($\mathrm{Dst}_{\min} \leq -250$ nT). After filtering for reliable main phase identification, 1643 storms were retained. In this filtered sample, the average main phase duration per class was 9.25 hours for moderate storms, 11.30 hours for intense storms and 13.75 hours for super-storms. After also requiring OMNIWeb solar wind data availability, 990 storms remained for examining the relationship between storm intensity and solar wind derived parameters over the main phase. The strongest correlations with $|\mathrm{Dst}_{\min}|$ were obtained for peak dawn-dusk electric field ($\mathrm{E}_{\mathrm{yp}}$) with $\mathrm{R} = 0.836$, peak southward IMF ($\mathrm{B}_{\mathrm{zp}}$) with $\mathrm{R} = -0.811$, and integrated electric field ($\mathrm{E}_{\mathrm{yi}}$) with $\mathrm{R} = 0.759$. These results confirm the importance of the solar wind electric field and southward interplanetary magnetic field in geomagnetic storm development. The main contribution of the application is its adjustable detection framework, which allows users to explore how parameter choices affect storm detection, main phase identification, and solar wind association results. All outputs are downloadable for further user-defined analysis. The application is available at: \boldurl{https://geomagnetic-storm-mainphase-analysis.streamlit.app/}
\end{abstract}

\section{Introduction}
\label{sec:introduction}

Geomagnetic storms are major disturbances of the terrestrial magnetosphere driven by enhanced solar wind energy input. The primary physical mechanism for this energy transfer is magnetic reconnection between the interplanetary magnetic field and the Earth's magnetic field, whose efficiency increases under sustained southward IMF conditions. Prolonged southward orientation of the interplanetary magnetic field has been empirically recognized as a primary solar wind driver of storm development \citep{gonzalez1987}. However, \citet{daglis2003} showed that the geoeffectiveness of this driver is also influenced by internal magnetospheric conditions, such as plasma sheet density, which can modify the resulting storm intensity.

Storm intensity is commonly measured by the Disturbance Storm Time index (Dst), which reflects variations in the horizontal component of the geomagnetic field measured at ground-based stations relative to quiet-time conditions. During the storm main phase, defined as the interval during which Dst decreases toward its minimum value, enhanced injection of energetic ions into the inner magnetosphere strengthens the ring current, producing the characteristic negative depression in Dst. Therefore, the minimum Dst value is commonly used as an indicator of storm intensity.

\begin{figure}[H]
\centering
\includegraphics[width=0.88\linewidth]{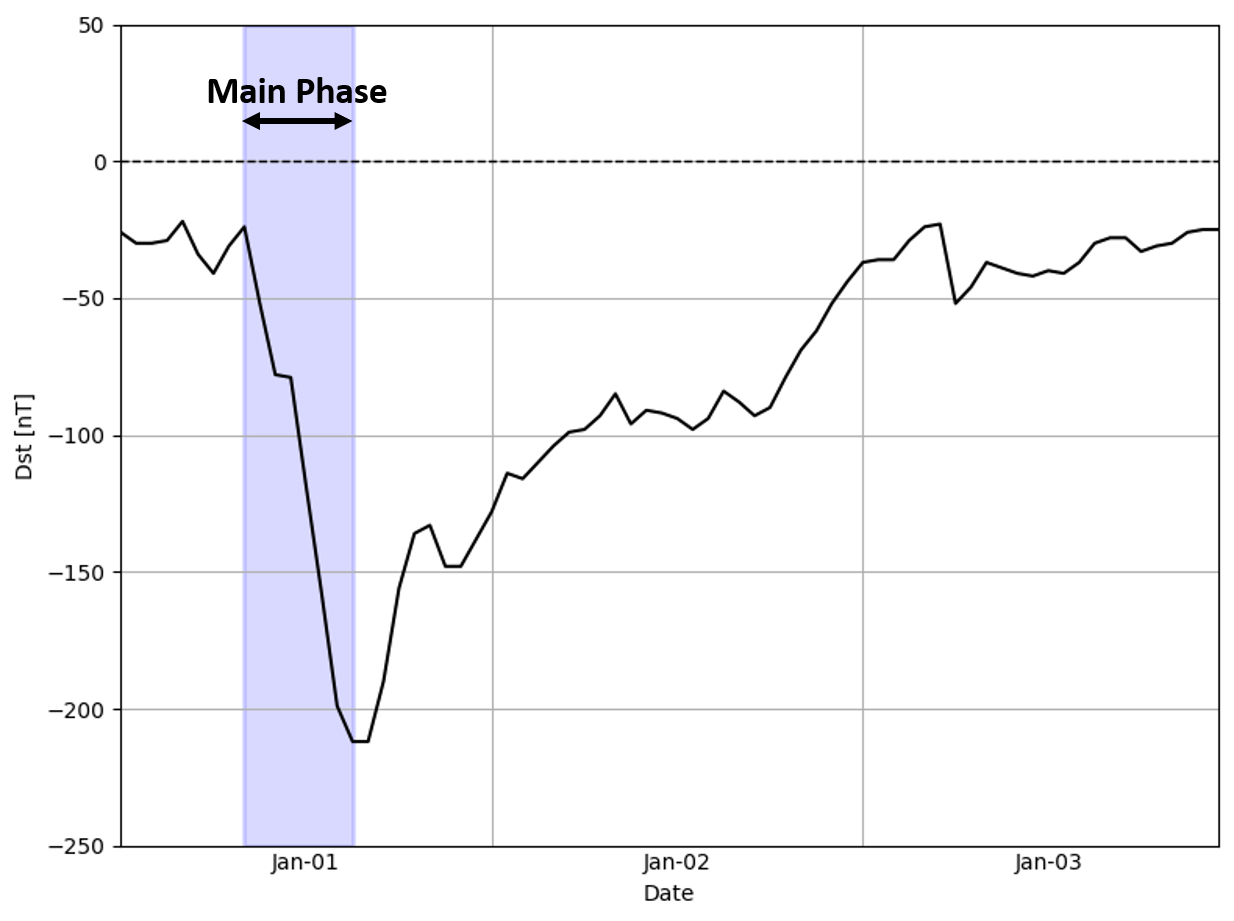}
\caption{Dst profile of an intense geomagnetic storm that occurred in 2025, reaching its minimum Dst = $-212$ nT on 1 January.}
\label{fig:jan2025}
\end{figure}

Figure~\ref{fig:jan2025} shows the Dst profile of an intense geomagnetic storm that occurred in January 2025, using hourly OMNIWeb values. The event provides an example of a storm with a clear main phase evolution, characterized by a sustained decrease in Dst toward a minimum value of $-212$ nT at 15:00 UT on 1 January 2025.

\begin{figure}[H]
\centering
\begin{tabular}{cc}
\includegraphics[width=0.46\linewidth]{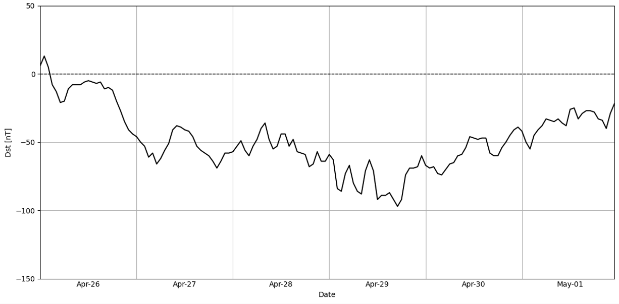} &
\includegraphics[width=0.46\linewidth]{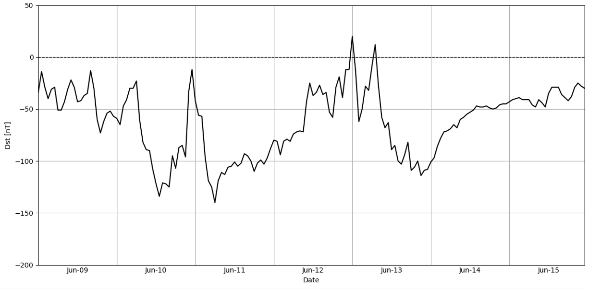} \\
\includegraphics[width=0.46\linewidth]{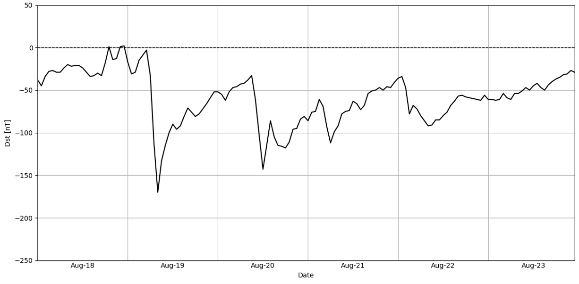} &
\includegraphics[width=0.46\linewidth]{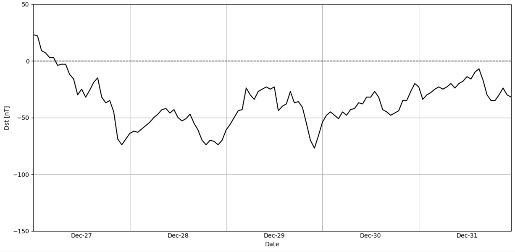}
\end{tabular}
\caption{Dst profiles of geomagnetic storms that occurred in 1991, during the solar maximum phase of Solar Cycle 22.}
\label{fig:complex1991}
\end{figure}

In this study, the analysis is limited to solar wind parameters as storm drivers and does not include internal magnetospheric conditions. It also does not attempt to identify the solar source or the specific interplanetary driver structures of the detected storms. However, \citet{zhang2007} noted that the storm-driving component, which possesses a prolonged southward orientation of the interplanetary magnetic field, may be an ICME sheath, multiple interacting ICMEs, a corotating interaction region (CIR), or a combination of these structures. Such complex driving conditions may lead to highly irregular Dst profiles. Figure~\ref{fig:complex1991} shows some examples of more complicated Dst profiles with multiple depressions.

This study's main objective is to give answers to the following questions:
\begin{itemize}
    \item When should a separate Dst depression be classified as a distinct storm?
    \item How is the main phase onset time determined?
    \item What is the relationship between storm intensity and solar wind?
\end{itemize}

It presents a web application for retrospective detection of geomagnetic storms from hourly Dst data and for identification of their main phases. The application allows users to adjust algorithm parameters, so that ambiguous events can be examined interactively rather than being classified only through one fixed set of assumptions. After storm detection and main phase identification, the application computes solar wind derived parameters during the main phase and examines their relationship with storm intensity.

\section{Data set}
\label{sec:data-set}

The analysis uses hourly averaged OMNI data obtained through NASA's OMNIWeb Data Explorer (\boldurl{https://omniweb.gsfc.nasa.gov/form/dx1.html}) \citep{papitashvili2020}, covering the period from 1964 through May 2026. The data used by the application are hourly values of the Disturbance Storm Time index (Dst), interplanetary magnetic field (IMF) magnitude, the GSM z-component of the interplanetary magnetic field ($\mathrm{B}_{\mathrm{z}}$), solar wind speed ($\mathrm{V}$) and dawn-dusk solar wind electric field ($\mathrm{E}_{\mathrm{y}}$). Dst, IMF, and $\mathrm{B}_{\mathrm{z}}$ are measured in nT, and $\mathrm{V}$ is measured in km/s. $\mathrm{E}_{\mathrm{y}}$ is derived from the product $-\mathrm{V}\mathrm{B}_{\mathrm{z}}$ and is expressed in mV/m as $\mathrm{E}_{\mathrm{y}} = -\mathrm{V}\,\mathrm{B}_{\mathrm{z}} \times 10^{-3}$.

\section{Algorithm and outputs}
\label{sec:algorithm}

The algorithm operates in three stages. First, geomagnetic storms are detected over the time period that the user has set. Second, the detected events are filtered, and their main phases are identified. Third, derived solar wind parameters are extracted for the filtered storms with data-complete main phases.

\subsection{Stage 1: Storm detection}
\label{sec:stage1}

The first stage identifies geomagnetic storms based on their Dst minima. At this stage, the program is concerned only with the minimum Dst depression of each storm and the time at which this minimum occurs.

\minititle{Stage 1 algorithm parameters:}
\textbf{EPISODE\_LEVEL} defines the Dst threshold used to identify continuous disturbed intervals. A disturbed episode begins when Dst becomes equal to or lower than EPISODE\_LEVEL and continues as long as consecutive hourly Dst values remain at or below this threshold. Thus, EPISODE\_LEVEL acts as the initial, broad criterion for identifying magnetically disturbed periods. The default value is {\bfseries\boldmath $-20$ nT}.

\textbf{STORM\_LEVEL} defines the Dst threshold used to identify storm-level minima within each disturbed episode. After a disturbed episode has been detected, the algorithm searches within that episode for candidate storm minima. A \textbf{candidate storm minimum} is defined as a Dst value that is equal to or lower than STORM\_LEVEL and is also a local minimum within the surrounding time window defined by LOCAL\_MIN\_RADIUS\_HOURS. The default value is {\bfseries\boldmath $-50$ nT}, which is commonly used as the threshold for identifying moderate geomagnetic storms.

\textbf{LOCAL\_MIN\_RADIUS\_HOURS} defines the time window used to test whether a storm-level Dst value is a local minimum. For each Dst value that is equal to or lower than STORM\_LEVEL, the algorithm checks whether it is the lowest Dst value within a surrounding time window of $\pm$LOCAL\_MIN\_RADIUS\_HOURS. If it is, the point is retained as a candidate storm minimum. The default value is \textbf{3 hours}.

\textbf{MINIMA\_SPLIT\_FACTOR} defines the \textbf{minimum-splitting criterion}, which controls whether multiple candidate storm minima within the same disturbed episode are treated as separate storm events. When two candidate storm minima occur within the same episode, the algorithm examines the maximum Dst value between them, referred to here as the inter-minimum peak. The inter-minimum peak height is defined as
\begin{equation}
\text{\textbf{peak height}} = \text{inter-minimum peak} - \text{shallower minimum}.
\end{equation}
This height is then compared with a separation threshold, defined as a fraction of the absolute depth of the deepest minimum:
\begin{equation}
\text{\textbf{separation threshold}} = -\text{deeper minimum} \times \mathrm{MINIMA\_SPLIT\_FACTOR}.
\end{equation}

If the peak height is greater than or equal to the separation threshold, the two minima are treated as separate storm events. Otherwise, they are considered part of the same storm structure, and the deeper minimum is retained as the representative minimum Dst of the storm. The default value for MINIMA\_SPLIT\_FACTOR is \textbf{0.4}.

If more than two candidate storm minima occur within the same episode, they are evaluated sequentially in chronological order. Each new candidate minimum is compared with the most recently retained minimum. If the separation criterion is satisfied, the new minimum is retained as a separate storm event. If not, the two minima are merged and only the deeper minimum is retained before the algorithm proceeds to the next candidate minimum.

\textbf{STORM\_LIMIT} imposes a lower bound on accepted Dst minima. It allows the user to restrict the analysis to a selected storm-intensity range and to exclude events outside the desired limits.

\textbf{Note:} To select a specific storm-intensity range, STORM\_LEVEL can be used as the upper threshold and STORM\_LIMIT as the lower threshold. For example, setting STORM\_LEVEL = $-50$ nT and STORM\_LIMIT = $-99$ nT restricts the analysis to storms with $-99~\mathrm{nT} \leq \mathrm{Dst}_{\min} \leq -50~\mathrm{nT}$ (moderate storms).

\paragraph{Outputs of Stage 1.}
Stage 1 produces the catalogue of all detected storms. The main table lists each detected storm together with its time of minimum Dst ($\mathrm{t}_{\min}$) and the corresponding minimum Dst value ($\mathrm{Dst}_{\min}$). Additional Stage 1 tables provide the characteristics of the disturbed episodes from which the storms were identified, including the episode start and end times, duration, and the number of detected storms associated with each episode. Multi-storm episodes are also reported separately, together with the corresponding storm minima identified within the same continuous Dst depression.

\subsection{Stage 2: Main phase identification}
\label{sec:stage2}

In the second stage, the program identifies the main phase of each storm detected in Stage 1. Because the end of the main phase is defined by the minimum Dst time, $\mathrm{t}_{\min}$, Stage 2 focuses on estimating the corresponding onset time of the main phase, $\mathrm{t}_{\mathrm{start}}$.

For each detected storm, the algorithm searches backward in time from $\mathrm{t}_{\min}$. This search uses a reference Dst value, referred to as the \textbf{mark}. Once the backward search reaches a Dst value equal to or higher than the mark, the first local Dst maximum encountered is assigned as $\mathrm{Dst}_{\mathrm{start}}$, and its timestamp is assigned as $\mathrm{t}_{\mathrm{start}}$. The main phase duration is then calculated as the time difference between $\mathrm{t}_{\mathrm{start}}$ and $\mathrm{t}_{\min}$.

The Stage 1 storm detection parameters are not reapplied during this stage. Stage 2 uses only the storm minima and their timestamps previously identified during storm detection. Consequently, the determination of $\mathrm{t}_{\mathrm{start}}$ is not constrained by the temporal limits of the original disturbed episode. Instead, the algorithm searches backward through the available Dst record until both the mark condition and the local maximum criterion are satisfied. If no valid $\mathrm{t}_{\mathrm{start}}$ can be identified, the storm is excluded from the filtered storm sample. In this study, filtered storms are defined as detected storms for which the main phase can be reliably identified.

\minititle{Stage 2 algorithm parameters:}
\textbf{STRONG\_STORM\_THRESHOLD} separates strong storms from weaker storms for the purpose of defining the mark that will be used for main phase onset identification. The default value is {\bfseries\boldmath $-100$ nT}. Storms with $\mathrm{Dst}_{\min}$ equal to or deeper than this threshold use the fixed mark value defined by STRONG\_MARK.

\textbf{STRONG\_MARK} is the mark used for strong storms. The default value is {\bfseries\boldmath $-50$ nT}. During the backward search from $\mathrm{t}_{\min}$, once Dst reaches or exceeds this value, the first local maximum encountered is assigned as $\mathrm{Dst}_{\mathrm{start}}$, and its timestamp is assigned as $\mathrm{t}_{\mathrm{start}}$.

For weaker storms, the mark is scaled according to storm intensity:
\begin{equation}
\text{\textbf{WEAK\_MARK}} = \mathrm{Dst}_{\min} \times \frac{\mathrm{STRONG\_MARK}}{\mathrm{STRONG\_STORM\_THRESHOLD}}.
\end{equation}
With the default values, this gives WEAK\_MARK $= \mathrm{Dst}_{\mathrm{min}} \times \frac{1}{2}$.

\textbf{LOCAL\_MAX\_RADIUS\_HOURS} defines the time window used to identify a local Dst maximum during the backward search for the storm onset. A point is accepted as a local maximum if it has the highest Dst value within a surrounding time window of $\pm$LOCAL\_MAX\_RADIUS\_HOURS. The default value is \textbf{3 hours}. Increasing this parameter makes the onset detection less sensitive to short-term Dst fluctuations. However, if the value is too large, an earlier local maximum may be selected when pre-storm variations are present, resulting in an overestimation of the main phase duration.

\minititle{Exclusion due to previous-storm overlap}
After $\mathrm{t}_{\mathrm{start}}$ has been estimated, the algorithm checks whether the Dst minimum of a previously detected storm occurs between $\mathrm{t}_{\mathrm{start}}$ and $\mathrm{t}_{\min}$ of the current storm. If such a previous minimum exists, the current storm is excluded from the filtered sample. This criterion prevents the algorithm from assigning a main phase interval that extends across an earlier storm minimum. In such cases, the estimated interval no longer represents the development of a single independent storm. This exclusion helps ensure that main phase duration and the subsequent analysis of relationships with solar wind parameters are calculated only for storms with a physically meaningful onset-to-minimum interval.

\minititle{Disturbance filtering}
After the previous-storm exclusion has been applied, the program optionally applies a disturbance filter. This filter is applied primarily to low-intensity storm events in which the Dst evolution before $\mathrm{Dst}_{\min}$ is irregular, containing multiple secondary dips. Such events are still geomagnetic storms, but they are separated from the filtered storm sample because their automatic $\mathrm{t}_{\mathrm{start}}$ determination is less reliable. In an idealized storm evolution, Dst decreases relatively continuously from the main phase onset toward $\mathrm{Dst}_{\min}$. In practice, however, many storms show irregular pre-minimum Dst variations, including secondary dips before the final minimum. For low-intensity storms, these variations can make the automatic onset detection particularly ambiguous and may produce disproportionately long main phase durations, for example several tens of hours for a moderate storm with $\mathrm{Dst}_{\min}$ near $-60$ nT. The disturbance filter therefore separates these ambiguous events from the filtered storm sample.

\textbf{DISTURBANCE\_FILTER} activates or deactivates this filtering step.

\textbf{DISTURB\_LEVEL} defines the intensity range to which the disturbance filter is applied. Only storms with $\mathrm{Dst}_{\min}$ shallower than DISTURB\_LEVEL are tested by the disturbance filter. With the default value of {\bfseries\boldmath $-100$ nT}, the filter is applied to moderate storms.

\textbf{DISTURB\_DIP\_COUNT} defines the minimum number of secondary Dst dips required for a storm to be classified as a disturbance. If the number of dips between $\mathrm{t}_{\mathrm{start}}$ and $\mathrm{t}_{\min}$ reaches this value, the event is excluded from the filtered storm sample and added to the disturbance category. The default value is \textbf{2}.

\textbf{DISTURB\_DIP\_RADIUS\_HOURS} defines the time window used to identify secondary Dst dips during disturbance filtering. Between $\mathrm{t}_{\mathrm{start}}$ and $\mathrm{t}_{\min}$, the algorithm checks whether each Dst value is the lowest value within a surrounding time window of $\pm$DISTURB\_DIP\_RADIUS\_HOURS. If it is, the point is counted as a secondary Dst dip. The default value is \textbf{3 hours}. A larger radius makes the filter less sensitive to short-term Dst fluctuations, reducing the number of detected secondary dips and making storms less likely to be classified as disturbances.

\paragraph{Outputs of Stage 2.}
Stage 2 produces the catalogue of filtered storms. For each retained storm, the output table includes the Stage 1 storm information, such as $\mathrm{t}_{\min}$ and $\mathrm{Dst}_{\min}$, together with the estimated main phase onset time, $\mathrm{t}_{\mathrm{start}}$, the corresponding $\mathrm{Dst}_{\mathrm{start}}$, and the main phase duration. Additional Stage 2 tables list storms excluded because of previous-storm overlap, as well as storms classified as disturbances, including their automatically estimated but unreliable $\mathrm{t}_{\mathrm{start}}$ values and main phase durations.

\subsection{Stage 3: Solar wind derived parameter computation}
\label{sec:stage3}

In Stage 3, the program uses the OMNIWeb solar wind data within the main phase intervals identified after Stage 2 filtering to compute the corresponding solar wind derived parameters for each storm. Before computing the final parameters, the program checks the OMNIWeb solar wind data for missing values during the main phase of each storm. If a single value is missing for a given parameter, it is replaced by linear interpolation using the neighboring valid values. Storms containing consecutive missing values in any solar wind parameter during their main phase are rejected from the final data-complete sample. This step ensures that the derived solar wind parameters are calculated only for storms with sufficiently complete input data.

When a single missing value occurred at either $\mathrm{t}_{\mathrm{start}}$ or $\mathrm{t}_{\min}$, it was interpolated using the valid hourly values one hour before and one hour after the missing value, even when one of these values lay outside the main phase interval. Values outside the interval were used only for interpolation and were not added to it.

For each data-complete storm, the derived solar wind parameters are computed over the identified main phase and include the IMF peak ($\mathrm{IMF}_{\mathrm{p}}$), $\mathrm{B}_{\mathrm{z}}$ peak ($\mathrm{B}_{\mathrm{zp}}$), solar wind speed peak ($\mathrm{V}_{\mathrm{p}}$), $\mathrm{E}_{\mathrm{y}}$ peak ($\mathrm{E}_{\mathrm{yp}}$) and integrated $\mathrm{E}_{\mathrm{y}}$ ($\mathrm{E}_{\mathrm{yi}}$), defined here as the sum of positive $\mathrm{E}_{\mathrm{y}}$ values within this interval. The $\mathrm{B}_{\mathrm{z}}$ peak is defined as the most negative $\mathrm{B}_{\mathrm{z}}$ value, whereas the other peak quantities correspond to the maximum values reached.

\paragraph{Outputs of Stage 3.}
Stage 3 produces the final catalogue of the data-complete storms. The main output table lists each Stage 2 storm that was not rejected due to incomplete solar wind data. In addition to the storm quantities already determined in the previous stages, namely $\mathrm{t}_{\min}$, $\mathrm{Dst}_{\min}$, $\mathrm{t}_{\mathrm{start}}$, $\mathrm{Dst}_{\mathrm{start}}$, and main phase duration, each storm is now listed together with its derived solar wind parameters. Additional Stage 3 output tables provide complementary information for the data-complete storms, including the time delays between $\mathrm{B}_{\mathrm{zp}}$, $\mathrm{E}_{\mathrm{yp}}$, and $\mathrm{Dst}_{\min}$, as well as the timestamps at which all peak parameters occur. The program also produces a list of storms for which a single missing value was replaced by interpolation, together with the replaced value, and a list of rejected storms, including the failed parameters and the number of missing values.

\subsection{Additional outputs}
\label{sec:more-outputs}

Additional outputs include Pearson correlation coefficients between the absolute storm intensity, $|\mathrm{Dst}_{\min}|$, and the derived parameters of the Stage 3 data-complete storms, together with scatter plots of $|\mathrm{Dst}_{\min}|$ versus $\mathrm{IMF}_{\mathrm{p}}$, $\mathrm{B}_{\mathrm{zp}}$, $\mathrm{V}_{\mathrm{p}}$, $\mathrm{E}_{\mathrm{yp}}$, $\mathrm{E}_{\mathrm{yi}}$, and main phase duration. The corresponding $\mathrm{p}$-values are calculated, and a correlation is considered statistically significant when $\mathrm{p}<0.05$. 

An optional parameter-specific analysis can also be performed directly on the Stage 2 filtered storm sample, independently of Stage 3. The user may select $\mathrm{IMF}$, $\mathrm{B}_{\mathrm{z}}$, or $\mathrm{V}$, after which data completeness is checked only for the selected parameter over each storm's main phase. The corresponding peak value is calculated for each retained storm, followed by its Pearson correlation with $|\mathrm{Dst}_{\min}|$. Consequently, a storm may be included in the analysis for one parameter even if missing values in another parameter caused it to be excluded from Stage 3. This provides a less restrictive correlation sample, which can be particularly useful when the selected storm sample is small.

Finally, the output tables from all stages are exported as downloadable CSV files.

\subsection{Examples}
\label{sec:examples}

\begin{figure}[H]
\centering
\includegraphics[width=0.95\linewidth]{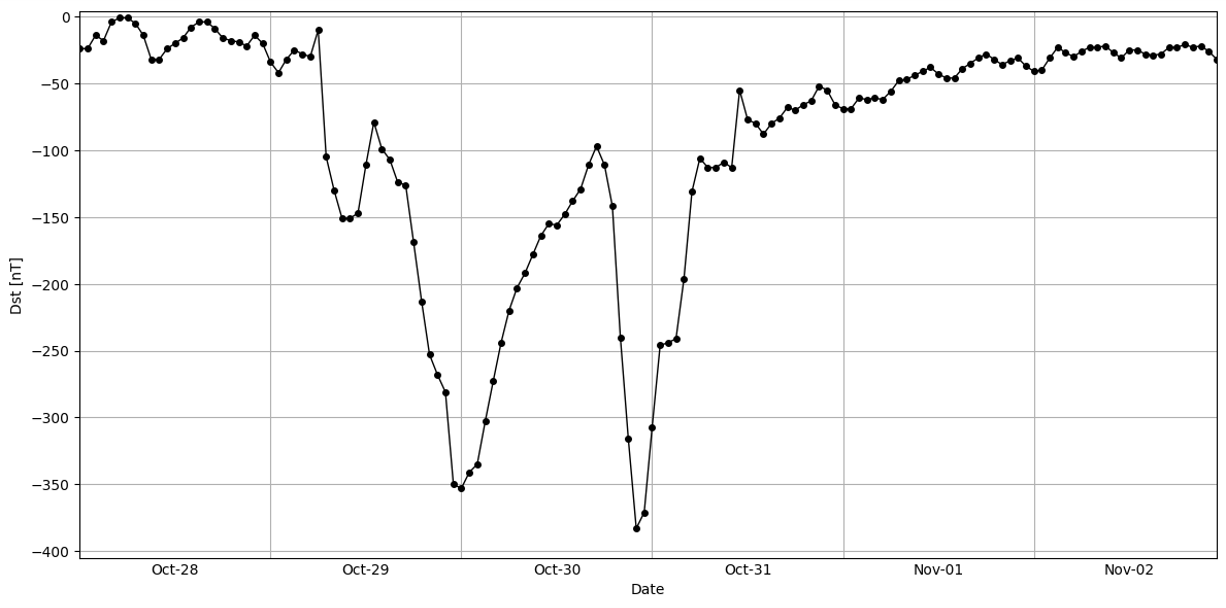}\\[0.3em]
\includegraphics[width=0.95\linewidth]{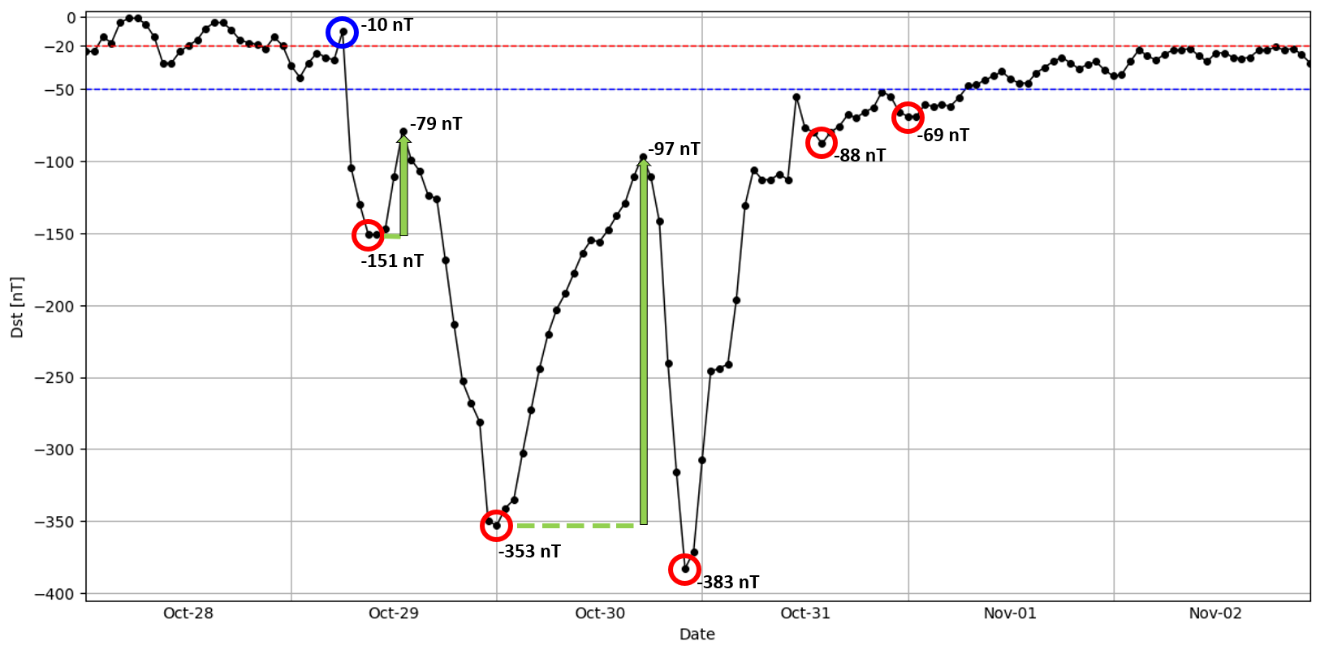}
\caption{Hourly Dst variation from 28 October to 2 November 2003. The upper panel shows the Dst profile during the selected interval. The lower panel illustrates how the algorithm evaluates the event. Red circles show the local Dst minima deeper than $-50$ nT. The green arrows indicate the inter-minimum peak heights. The blue circle shows the main phase onset of the first super-storm event with $\mathrm{Dst}_{\min} = -353$ nT. The red dashed line denotes the episode level, and the blue dashed line denotes the strong-storm Dst reference mark used for main phase onset identification.}
\label{fig:oct2003}
\end{figure}

\phantomsection
\label{ex:example1}
\paragraph{Example 1.}
In this example, the October 2003 super-storm sequence is used to illustrate how the algorithm operates across all three stages. Figure~\ref{fig:oct2003} shows the hourly Dst variation from 28 October to 2 November 2003 in the upper panel and how the algorithm would evaluate this event in the lower panel. Within the main disturbed episode, defined by EPISODE\_LEVEL = $-20$ nT and shown by the red dashed line, the algorithm searches for candidate storm minima. These are local Dst minima within a $\pm 3$ h window, set by LOCAL\_MIN\_RADIUS\_HOURS = 3 h, with values equal to or lower than STORM\_LEVEL = $-50$ nT. Five candidate minima are identified: $-151$ nT, $-353$ nT, $-383$ nT, $-88$ nT, and $-69$ nT, shown in red circles.

The minimum-splitting criterion was then applied sequentially to determine whether consecutive candidate minima should be retained as separate storm events or merged into the same storm structure. The first candidate minimum, Dst = $-151$ nT, was compared with the following deeper minimum, Dst = $-353$ nT. The inter-minimum peak between them was $-79$ nT. Therefore,
\begin{equation}
\mathrm{peak\ height} = -79 - (-151) = 72~\mathrm{nT}.
\end{equation}
This peak height is represented by the smaller green arrow in the figure. The corresponding separation threshold was
\begin{equation}
\mathrm{separation\ threshold} = -(-353) \times 0.4 = 141.2~\mathrm{nT}.
\end{equation}
Since the peak height was smaller than the separation threshold, the two minima were assigned to the same storm structure, and the deeper Dst = $-353$ nT minimum was retained.

The retained Dst = $-353$ nT minimum was then compared with the next candidate minimum, Dst = $-383$ nT. The inter-minimum peak between them was $-97$ nT. Therefore,
\begin{equation}
\mathrm{peak\ height} = -97 - (-353) = 256~\mathrm{nT}.
\end{equation}
This peak height is represented by the larger green arrow. The corresponding separation threshold was
\begin{equation}
\mathrm{separation\ threshold} = -(-383) \times 0.4 = 153.2~\mathrm{nT}.
\end{equation}
In this case, the peak height exceeded the separation threshold. Therefore, the Dst = $-383$ nT minimum was retained as a separate storm event.

The two later candidate minima, Dst = $-88$ nT and Dst = $-69$ nT, were then compared sequentially with the currently retained Dst = $-383$ nT minimum. In both cases, the inter-minimum peak was insufficient to satisfy the minimum-splitting criterion, so these later minima were not retained as separate storm events.

As a result, Stage 1 detected two storms within the same disturbed episode: one with $\mathrm{Dst}_{\min} = -353$ nT at 30 October 2003, 00:00 UTC, and another with $\mathrm{Dst}_{\min} = -383$ nT at 30 October 2003, 22:00 UTC.

Stage 2 was then applied to the two detected storms in order to identify their onset times and assess whether their main phase intervals were suitable for further analysis. Both detected storms had $\mathrm{Dst}_{\min}$ values deeper than STRONG\_STORM\_THRESHOLD = $-100$ nT. Therefore, for both storms, the strong-storm Dst reference mark STRONG\_MARK = $-50$ nT was used, shown by the blue dashed line. During the backward search from each storm minimum, after crossing this mark, the algorithm selected the first local Dst maximum within a $\pm 3$ h window, set by LOCAL\_MAX\_RADIUS\_HOURS = 3 h.

For the first storm, with $\mathrm{Dst}_{\min} = -353$ nT, the algorithm identified $\mathrm{Dst}_{\mathrm{start}}$ at 29 October 2003 06:00 UTC, when Dst = $-10$ nT, shown in the blue circle. The resulting main phase duration was 18 hours, and the storm was retained in the filtered storm sample.

For the second storm, with $\mathrm{Dst}_{\min} = -383$ nT, the backward search led to the same $\mathrm{Dst}_{\mathrm{start}}$ point, so the $\mathrm{Dst}_{\min}$ of the first storm occurred between this $\mathrm{Dst}_{\mathrm{start}}$ time and the $\mathrm{Dst}_{\min}$ of the second storm. Therefore, the estimated main phase of the second storm overlapped the previous storm and was not considered independent. As a result, the second storm was excluded from the filtered storm sample.

In Stage 3, the storm retained after Stage 2 was tested for OMNI data completeness over its main phase, from 29 October 2003 at 06:00 UTC to 30 October 2003 at 00:00 UTC. Although the Dst main phase was well defined, the V and $\mathrm{E}_{\mathrm{y}}$ parameters contained consecutive missing values throughout the interval. Because the algorithm only replaces isolated internal missing values, this storm was rejected from the final data-complete sample.

Thus, Stage 1 detected two storm events within the disturbed episode, with $\mathrm{Dst}_{\min} = -353$ nT and $\mathrm{Dst}_{\min} = -383$ nT. Stage 2 retained only the first storm in the filtered storm sample, while Stage 3 rejected it from the final data-complete sample.

\begin{figure}[H]
\centering
\includegraphics[width=0.95\linewidth]{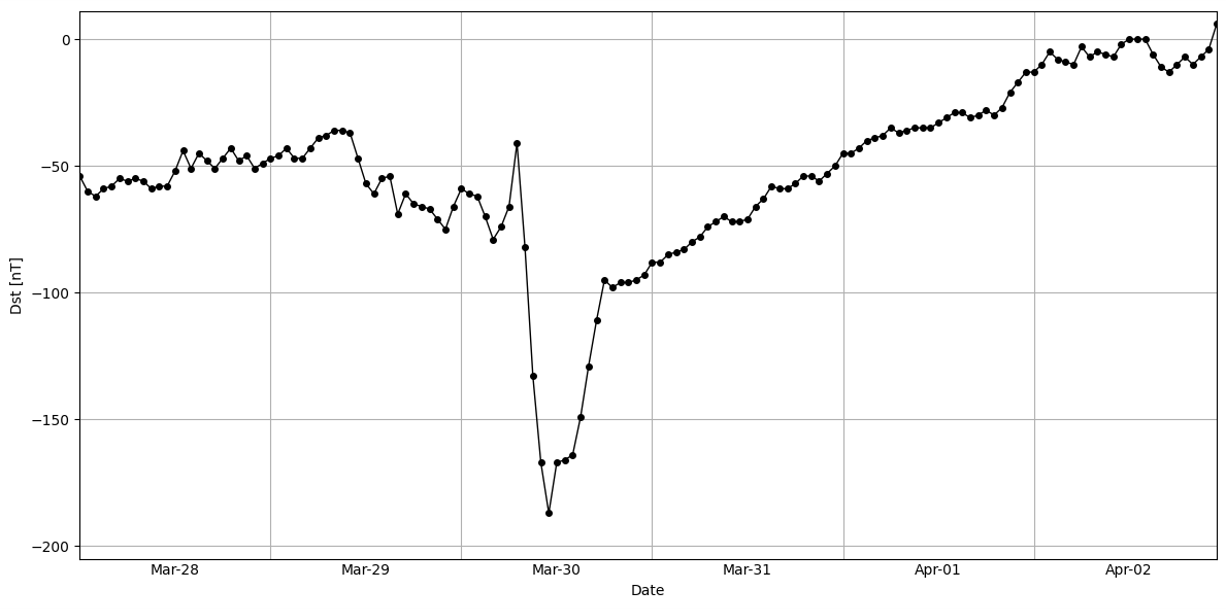}\\[0.3em]
\includegraphics[width=0.95\linewidth]{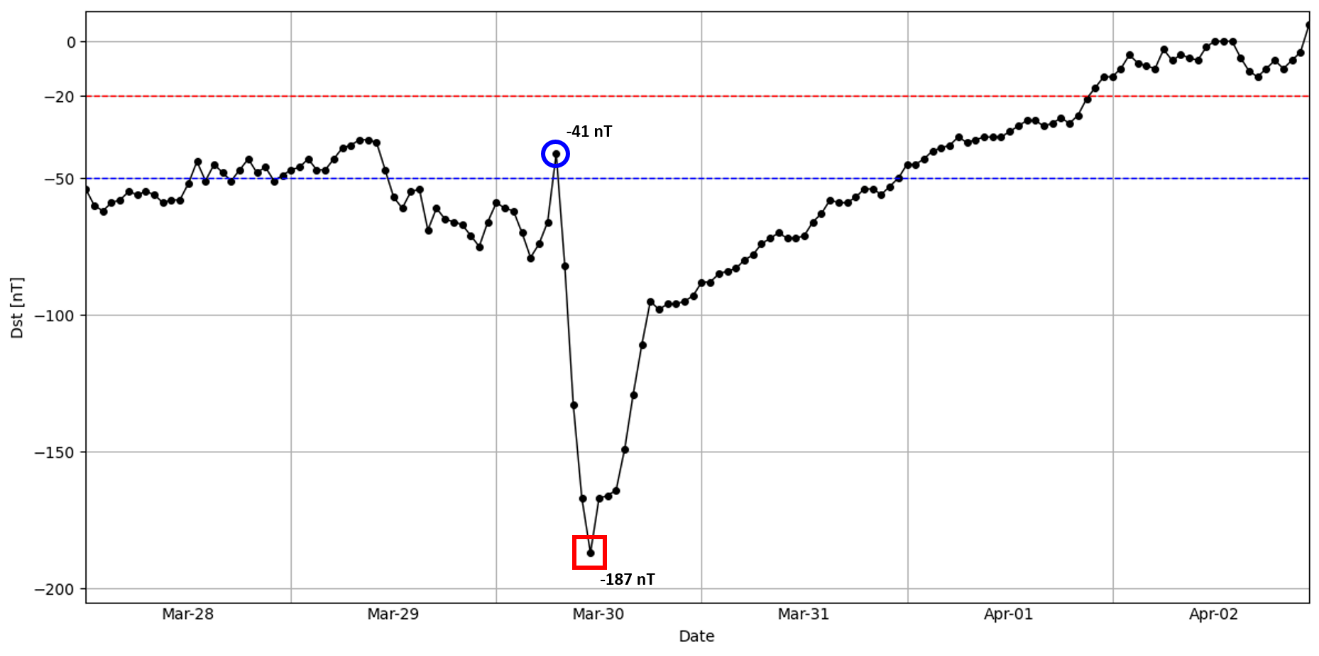}
\caption{Hourly Dst variation from 28 March to 2 April 1990. The upper panel shows the Dst profile during the selected interval. The lower panel highlights the detected storm minimum, shown in the red square, and the main phase onset, shown in the blue circle. The red dashed line denotes the episode level and the blue dashed line denotes the strong-storm Dst reference mark used for main phase onset identification.}
\label{fig:mar1990}
\end{figure}

\FloatBarrier
\newpage
\paragraph{Example 2.}
In this example, the importance of the reference mark in the main phase onset identification procedure is demonstrated. Figure~\ref{fig:mar1990} shows the hourly Dst variation from 28 March to 2 April 1990 in the upper panel and how the algorithm would evaluate this event with focus on the Stage 2 strong-storm Dst reference mark. Within the detected episode, defined by EPISODE\_LEVEL = $-20$ nT and shown by the red dashed line, the algorithm searches for local Dst minima using a $\pm 3$ h window, set by LOCAL\_MIN\_RADIUS\_HOURS = 3 h. The deepest minimum occurs on 30 March 1990 at 11:00 UTC, where Dst reaches $-187$ nT, shown in the red square. Although additional local minima deeper than STORM\_LEVEL = $-50$ nT occur earlier within the same episode, they are not classified as separate storm events because they do not satisfy the minimum-splitting criterion relative to the dominant $-187$ nT minimum.

As a result, Stage 1 detected one storm during this time period with $\mathrm{Dst}_{\min} = -187$ nT at 30 March 1990, 11:00 UTC.

In Stage 2, the main phase onset is determined for the detected storm. Since the storm is stronger than the STRONG\_STORM\_THRESHOLD = $-100$ nT, the strong-storm Dst reference mark, STRONG\_MARK = $-50$ nT, is used, shown by the blue dashed line. Searching backward from the storm minimum, the algorithm identifies the first point at which Dst crosses this mark. This occurs on 30 March 1990 at 07:00 UTC, where Dst = $-41$ nT. Since this point is also a local maximum within a $\pm 3$ h window, set by LOCAL\_MAX\_RADIUS\_HOURS = 3 h, it is accepted as the main phase onset, shown in the blue circle.

Thus, the storm main phase extends from 30 March 1990 at 07:00 UTC to 30 March 1990 at 11:00 UTC, with a total duration of 4 h.

This example illustrates the importance of the strong-storm Dst reference mark, STRONG\_MARK. A disturbed interval is present before the rapid storm development. If the backward search had instead relied on a fixed quiet-time threshold, the inferred onset would have shifted to a much earlier part of the disturbed interval, resulting in an overestimated main phase duration that would not represent the actual storm build-up.

\begin{figure}[H]
\centering
\includegraphics[width=0.95\linewidth]{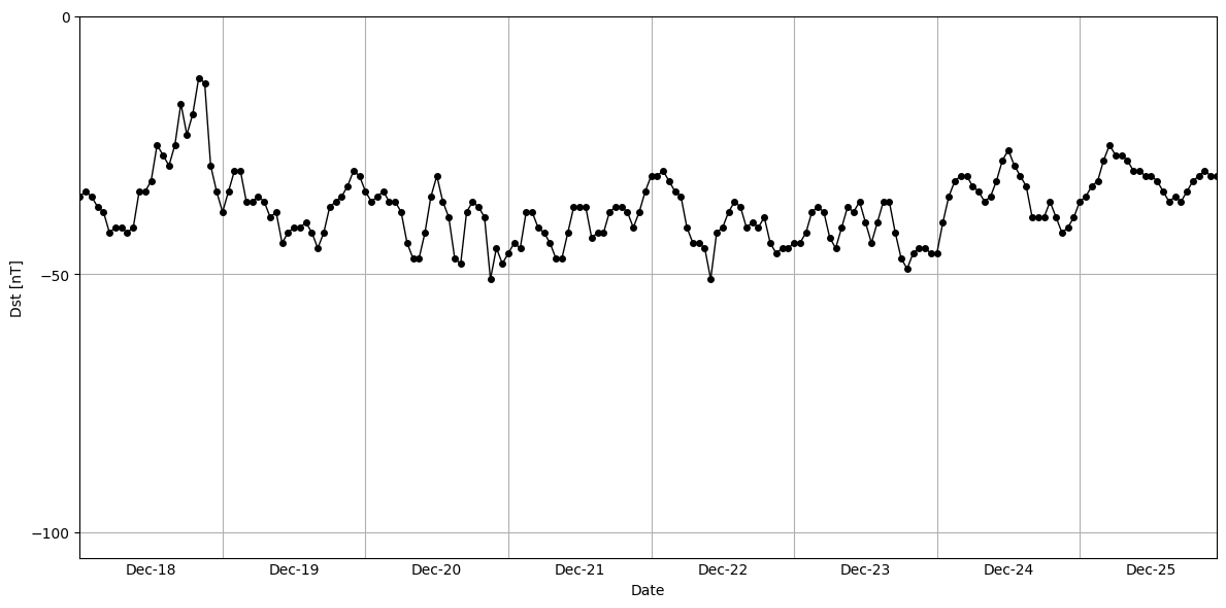}\\[0.3em]
\includegraphics[width=0.95\linewidth]{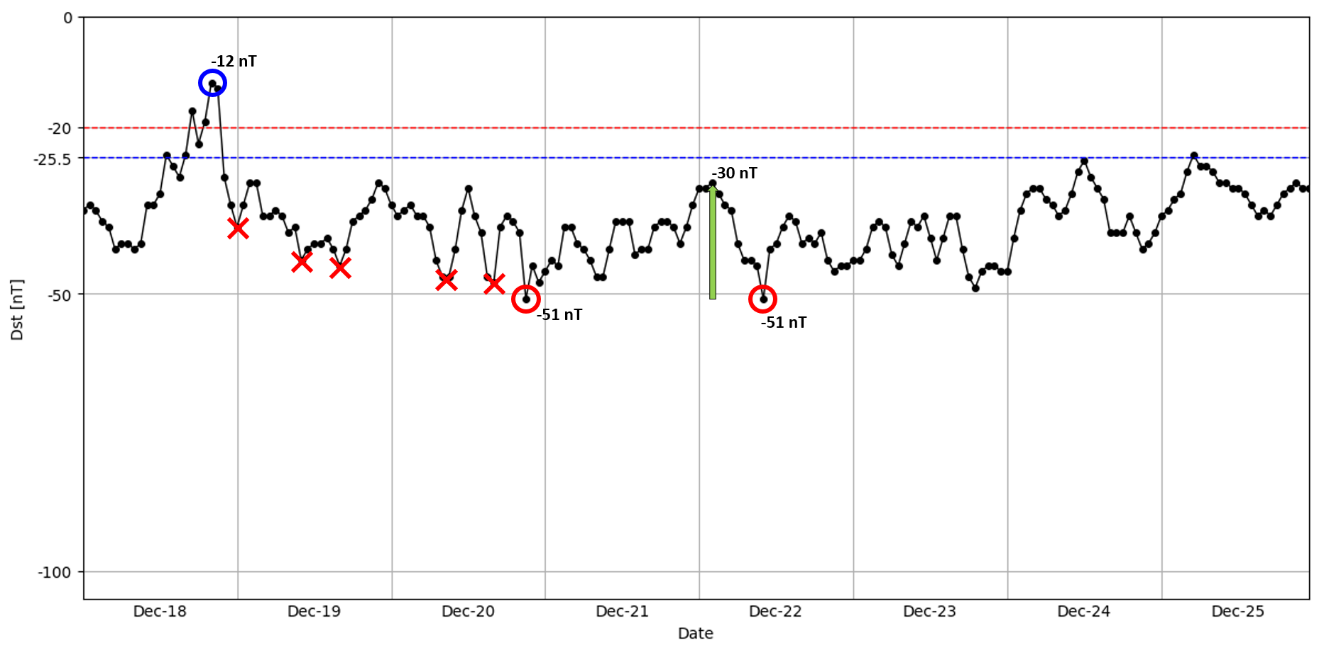}
\caption{Hourly Dst variation from 18 to 25 December 2006. The upper panel shows the Dst profile during the selected interval. The lower panel illustrates how the algorithm evaluates the event. Red circles show the local Dst minima deeper than $-50$ nT. The green arrow indicates the inter-minimum peak height. The blue circle shows the identified main phase onset of the first weak storm of $\mathrm{Dst}_{\min} = -51$ nT before its rejection by the disturbance filter. The red dashed line denotes the episode level and the blue dashed line denotes the weak-storm Dst reference mark used for main phase onset identification.}
\label{fig:dec2006}
\end{figure}

\FloatBarrier
\newpage
\paragraph{Example 3.}
This example illustrates the operation of the weak-storm Dst reference mark and the disturbance filter. Figure~\ref{fig:dec2006} shows the hourly Dst variation from 18 to 25 December 2006 in the upper panel and how the algorithm would evaluate this event in the lower panel. Within the main detected episode, defined by EPISODE\_LEVEL = $-20$ nT and shown by the red dashed line, the algorithm finds two local Dst minima deeper than STORM\_LEVEL = $-50$ nT: one on 20 December 2006 at 21:00 UTC, where Dst = $-51$ nT, and another on 22 December 2006 at 10:00 UTC, where Dst = $-51$ nT.

The minimum-splitting criterion is then applied to determine whether these two candidate minima should be treated as separate storm events. The inter-minimum peak between them occurs on 22 December 2006 at 02:00 UTC, where Dst = $-30$ nT. Therefore, the peak height is
\begin{equation}
\mathrm{peak\ height} = -30 - (-51) = 21~\mathrm{nT}.
\end{equation}
This peak height is represented by the green arrow. The corresponding separation threshold was
\begin{equation}
\mathrm{separation\ threshold} = -(-51) \times 0.4 = 20.4~\mathrm{nT}.
\end{equation}
Since the peak height exceeds the separation threshold, Stage 1 classifies the two minima as separate storm events.

In Stage 2, both detected storms are treated as weak storms because their Dst minima are not deeper than STRONG\_STORM\_THRESHOLD = $-100$ nT. Therefore, the Dst reference mark is scaled from the storm intensity, giving a value of
\begin{equation}
\text{\textbf{WEAK\_MARK}} = \mathrm{Dst}_{\min} \times \frac{\mathrm{STRONG\_MARK}}{\mathrm{STRONG\_STORM\_THRESHOLD}} = -51 \times \frac{-50}{-100} = -25.5~\mathrm{nT}
\end{equation}
for both storms, shown by the blue dashed line. For the first storm, the backward search identifies the main phase onset on 18 December 2006 at 20:00 UTC, where Dst = $-12$ nT, shown in the blue circle. However, the interval between this onset and the storm minimum contains multiple secondary Dst dips, indicated by red crosses, defined as being the lowest value within the surrounding $\pm$DISTURB\_DIP\_RADIUS\_HOURS = 3 h window. Since the number of Dst dips is at least DISTURB\_DIP\_COUNT = 2, the event is classified as a disturbance and is rejected from the filtered storm sample. For the second storm, the backward search identified the same main phase onset time. As a result, the $\mathrm{Dst}_{\min}$ of the first detected storm occurred between this onset time and the $\mathrm{Dst}_{\min}$ time of the second storm. Therefore, the estimated main phase interval of the second storm overlapped the previous storm and was not considered independent. Consequently, the second storm was also excluded from the filtered storm sample.

Thus, although Stage 1 detects two storms with $\mathrm{Dst}_{\min} = -51$ nT, Stage 2 rejects both. Consequently, no storm from this interval is retained in the filtered storm sample.

\FloatBarrier
\newpage
\section{Results}
\label{sec:results}

The results presented in this work are based on data covering the period from 1964 to May 2026 and were obtained using the default algorithm parameters of the application. The total storms detected with $\mathrm{Dst}_{\min} \leq -50$ nT were \textbf{1796}. Table~\ref{tab:detected} shows the detected storm count per class (moderate storms: $-100 < \mathrm{Dst}_{\min} \leq -50$ nT, intense storms: $-250 < \mathrm{Dst}_{\min} \leq -100$ nT, super-storms: $\mathrm{Dst}_{\min} \leq -250$ nT).

\begin{table}[htbp]
\centering
\setlength{\tabcolsep}{12pt}
\caption{Detected storm count per class.}
\label{tab:detected}
\begin{tabular}{lr}
\toprule
\textbf{Class} & \textbf{Detected storms} \\
\midrule
Moderate ($-100 < \mathrm{Dst}_{\min} \leq -50$) & 1422 \\
Intense ($-250 < \mathrm{Dst}_{\min} \leq -100$) & 347 \\
Super-storm ($\mathrm{Dst}_{\min} \leq -250$) & 27 \\
\bottomrule
\end{tabular}
\end{table}

After filtering for reliable main phase identification, \textbf{1643} storms were retained, with an average main phase duration of {\bfseries\boldmath $9.73 \pm 0.13$ hours}. Table~\ref{tab:filtered} shows the filtered storm count and the average main phase duration per class.

\begin{table}[htbp]
\centering
\setlength{\tabcolsep}{12pt}
\caption{Filtered storm count and average main phase duration per class.}
\label{tab:filtered}
\begin{tabular}{lcr}
\toprule
\textbf{Class} & \textbf{Filtered storms} & \textbf{Avg. Mp duration (h)} \\
\midrule
Moderate & 1284 & $9.25 \pm 0.13$ \\
Intense & 335 & $11.30 \pm 0.34$ \\
Super-storm & 24 & $13.75 \pm 1.52$ \\
\bottomrule
\end{tabular}
\end{table}

After also requiring OMNIWeb solar wind data availability, \textbf{990} storms were available for examining the relationship between storm intensity and solar wind derived parameters during the main phase. Table~\ref{tab:correlation990} shows the correlations between $|\mathrm{Dst}_{\min}|$ and solar wind derived parameters, main phase duration for all data-complete storms. Among these 990 storms, 796 were moderate, 184 were intense and 10 were super-storms. The average time delay of the minimum Dst from $\mathrm{E}_{\mathrm{yp}}$ and $\mathrm{B}_{\mathrm{zp}}$ was $3.83 \pm 0.10$ h and $3.84 \pm 0.10$ h respectively. Figure~\ref{fig:scatterplots} shows scatter plots of $|\mathrm{Dst}_{\min}|$ against the parameters listed in Table~\ref{tab:correlation990}.

\begin{table}[htbp]
\centering
\setlength{\tabcolsep}{12pt}
\caption{$\mathrm{R}$ and p-values of $|\mathrm{Dst}_{\min}|$ and solar wind derived parameters, main phase duration ($\mathrm{N}=990$, $|\mathrm{Dst}_{\min}| \geq 50$ nT).}
\label{tab:correlation990}
\begin{tabular}{lcr}
\toprule
\textbf{Derived parameter} & {\bfseries\boldmath $\mathrm{R}$} & \textbf{p-value} \\
\midrule
$\mathrm{IMF}_{\mathrm{p}}$ & 0.723 & $3.67\times10^{-161}$ \\
\corrhighlight{$\mathrm{B}_{\mathrm{zp}}$} & \corrhighlight{-0.811} & $2.57\times10^{-232}$ \\
\corrhighlight{$\mathrm{E}_{\mathrm{yp}}$} & \corrhighlight{0.836} & $9.74\times10^{-260}$ \\
\corrhighlight{$\mathrm{E}_{\mathrm{yi}}$} & \corrhighlight{0.759} & $3.24\times10^{-186}$ \\
$\mathrm{V}_{\mathrm{p}}$ & 0.288 & $2.14\times10^{-20}$ \\
Mp duration & 0.115 & $3.00\times10^{-4}$ \\
\bottomrule
\end{tabular}
\end{table}

By setting \textbf{STORM\_LEVEL} to {\bfseries\boldmath $-200$ nT}, all storms with $|\mathrm{Dst}_{\min}| \geq 200$ nT were detected over the same time period (1964 to May 2026). A total of 59 storms were detected, 55 were retained after main phase filtering, and the final data-complete count was 31. Table~\ref{tab:correlation31} shows the correlations between $|\mathrm{Dst}_{\min}|$ and solar wind derived parameters, main phase duration for the data-complete storms with $|\mathrm{Dst}_{\min}| \geq 200$ nT.

\begin{table}[htbp]
\centering
\setlength{\tabcolsep}{12pt}
\caption{$\mathrm{R}$ and p-values of $|\mathrm{Dst}_{\min}|$ and solar wind derived parameters, main phase duration ($\mathrm{N}=31$, $|\mathrm{Dst}_{\min}| \geq 200$ nT).}
\label{tab:correlation31}
\begin{tabular}{lcr}
\toprule
\textbf{Derived parameter} & {\bfseries\boldmath $\mathrm{R}$} & \textbf{p-value} \\
\midrule
$\mathrm{IMF}_{\mathrm{p}}$ & 0.586 & $5.34\times10^{-4}$ \\
\corrhighlight{$\mathrm{B}_{\mathrm{zp}}$} & \corrhighlight{-0.761} & $6.63\times10^{-7}$ \\
\corrhighlight{$\mathrm{E}_{\mathrm{yp}}$} & \corrhighlight{0.584} & $5.61\times10^{-4}$ \\
\corrhighlight{$\mathrm{E}_{\mathrm{yi}}$} & \corrhighlight{0.642} & $9.77\times10^{-5}$ \\
$\mathrm{V}_{\mathrm{p}}$ & 0.049 & $7.93\times10^{-1}$ \\
Mp duration & -0.079 & $6.74\times10^{-1}$ \\
\bottomrule
\end{tabular}
\end{table}

Since this sample is relatively small, data completeness was also examined separately for each solar wind parameter. The data-complete storm set remained unchanged for $\mathrm{E}_{\mathrm{y}}$, while the corresponding set for $\mathrm{B}_{\mathrm{z}}$ included 4 additional storms. These 4 cases had available $\mathrm{B}_{\mathrm{z}}$ values but missing solar wind speed (V) values, so $\mathrm{E}_{\mathrm{y}}$ was also unavailable, since it is derived from the product of V and $\mathrm{B}_{\mathrm{z}}$. Using the parameter-specific data-complete sample for $\mathrm{B}_{\mathrm{z}}$, the $\mathrm{B}_{\mathrm{zp}}$ sample increased to $\mathrm{N} = 35$, yielding a correlation coefficient of {\bfseries\boldmath $\mathrm{R} = -0.668$} with $|\mathrm{Dst}_{\min}|$.

\section{Web application screenshots}
\label{sec:screenshots}

\begin{figure}[H]
\centering
\includegraphics[width=0.95\linewidth]{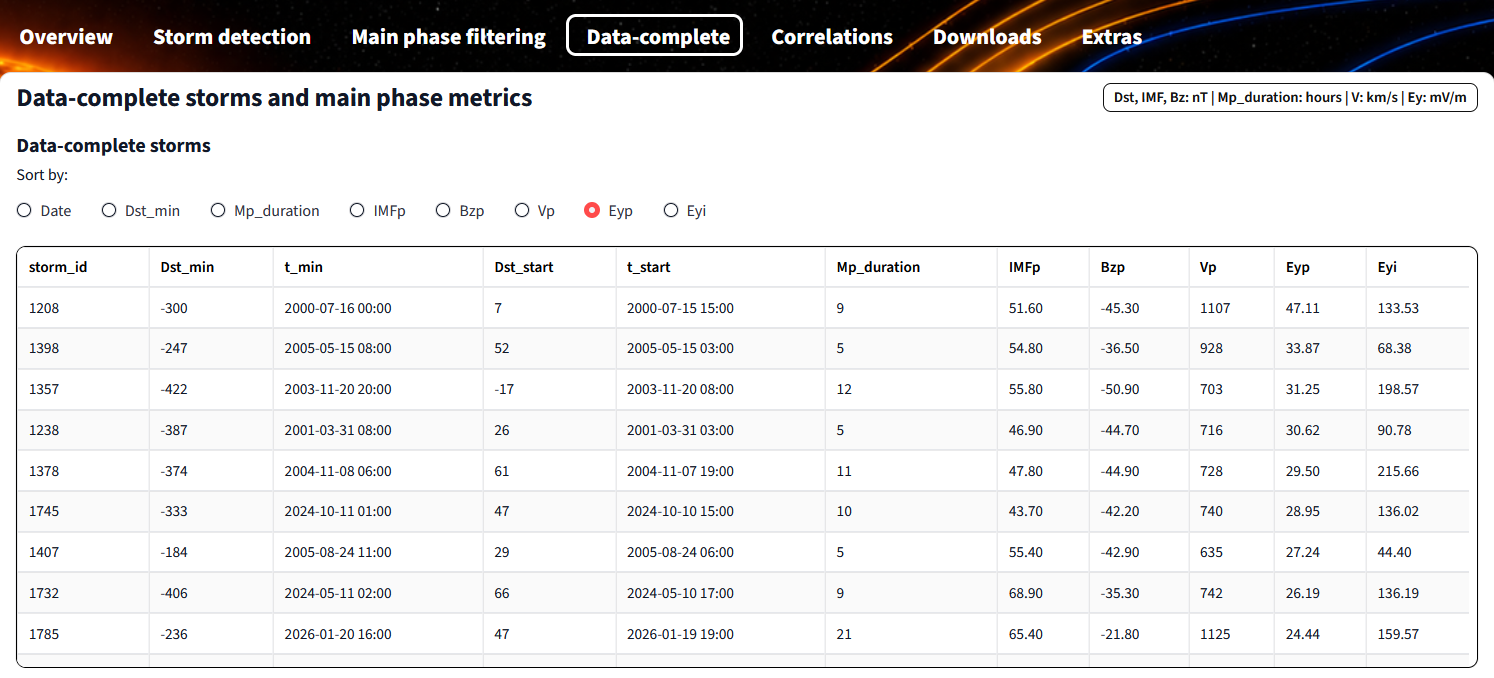}
\caption{Application screenshot of the data-complete storm catalogue, sorted from the strongest to the weakest peak dawn-dusk solar wind electric field ($\mathrm{E}_{\mathrm{yp}}$). Only the first nine storms are visible in this view.}
\label{fig:app-screenshot}
\end{figure}

\begin{figure}[H]
\centering
\setlength{\tabcolsep}{2pt}
\renewcommand{\arraystretch}{1.02}
\begin{tabular}{@{}cc@{}}
\includegraphics[width=0.48\linewidth]{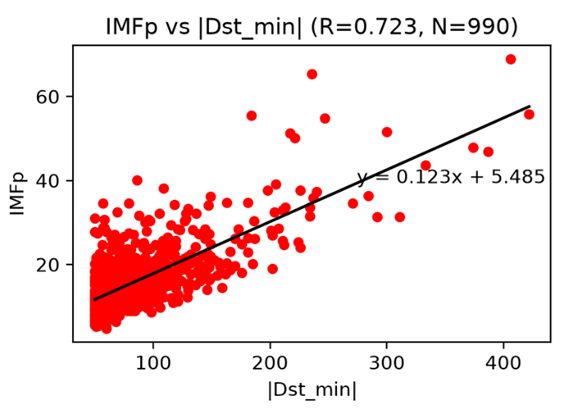} &
\includegraphics[width=0.48\linewidth]{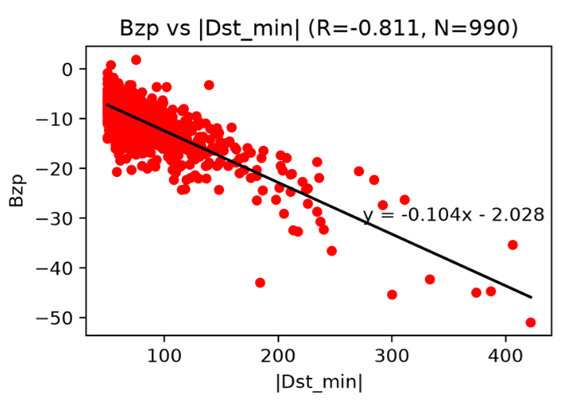} \\
\includegraphics[width=0.48\linewidth]{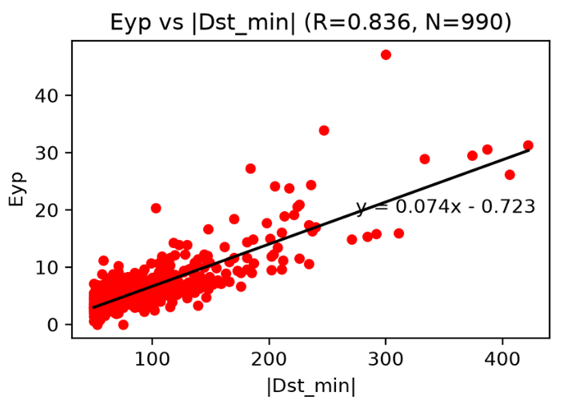} &
\includegraphics[width=0.48\linewidth]{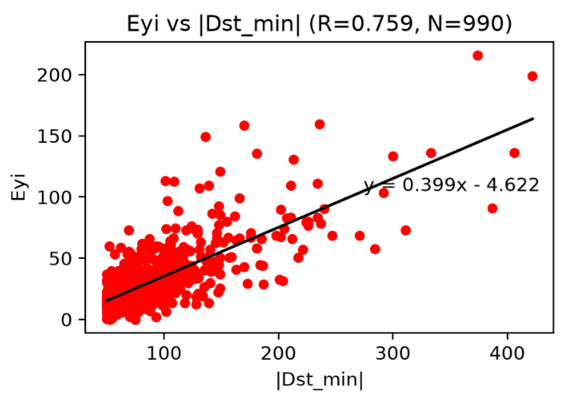} \\
\includegraphics[width=0.48\linewidth]{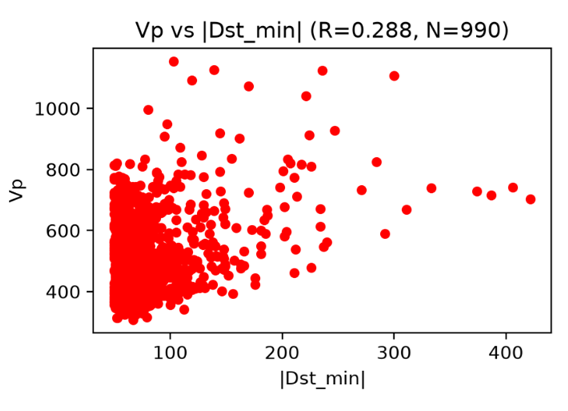} &
\includegraphics[width=0.48\linewidth]{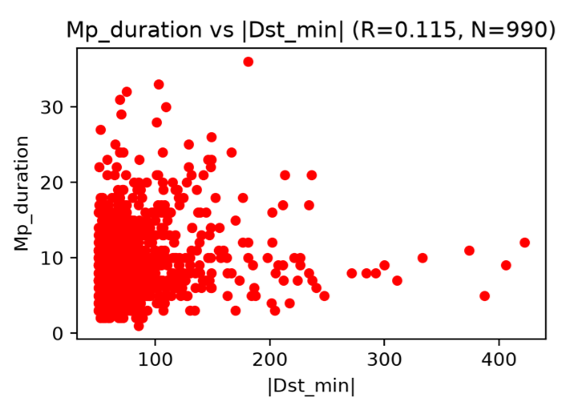}
\end{tabular}
\caption{Scatter plots of $|\mathrm{Dst}_{\min}|$ and derived parameters: $\mathrm{IMF}_{\mathrm{p}}$, $\mathrm{B}_{\mathrm{zp}}$, $\mathrm{E}_{\mathrm{yp}}$, $\mathrm{E}_{\mathrm{yi}}$, $\mathrm{V}_{\mathrm{p}}$, main phase duration. The solid lines indicate least-squares linear fits of $|\mathrm{Dst}_{\min}|$ vs. $\mathrm{IMF}_{\mathrm{p}}$, $\mathrm{B}_{\mathrm{zp}}$, $\mathrm{E}_{\mathrm{yp}}$ and $\mathrm{E}_{\mathrm{yi}}$.}
\label{fig:scatterplots}
\end{figure}

\section{Discussion}
\label{sec:discussion}

The main contribution of the present work lies in the development of a flexible and reproducible analysis framework. Whereas previous geomagnetic storm studies often rely on fixed thresholds or predefined event lists, the proposed application provides an adjustable framework in which users can modify the parameters controlling storm detection and main phase identification and examine how these choices affect the resulting outputs. In this sense, the application also serves as an organizational tool, as it collects and structures information on geomagnetic storms and their main phases. All generated outputs are downloadable and can be used for further user-defined analysis.

The results confirm the expected importance of southward IMF orientation and the solar wind electric field in geomagnetic storm development. This is consistent with the \citet{burton1975} Dst formulation, which links the ring current injection rate to the dawn-dusk component of the interplanetary electric field for southward IMF. The strongest associations with $|\mathrm{Dst}_{\min}|$ are obtained for $\mathrm{E}_{\mathrm{y}}$, $\mathrm{B}_{\mathrm{z}}$, and integrated $\mathrm{E}_{\mathrm{y}}$. This agrees with the physical interpretation that sustained southward IMF and enhanced dawn-dusk electric field are efficient drivers of storm intensification. More specifically, the correlation analysis of derived solar wind parameters vs. $|\mathrm{Dst}_{\min}|$ for the full data-complete sample of 990 storms with $|\mathrm{Dst}_{\min}| \geq 50$ nT showed that the strongest indicator of storm intensity is $\mathrm{E}_{\mathrm{yp}}$ ($\mathrm{R} = 0.836$), followed by $\mathrm{B}_{\mathrm{zp}}$ ($\mathrm{R} = -0.811$) and then by $\mathrm{E}_{\mathrm{yi}}$ ($\mathrm{R} = 0.759$). This suggests that, when all three storm classes are included in the analysis (moderate, intense, super-storm), the peak value of the dawn-dusk solar wind electric field during the main phase shows the strongest association with storm intensity. It is important to note that the sample is dominated by moderate storms ($-100 < \mathrm{Dst}_{\min} \leq -50$), which account for about 80\% of all events. Although $\mathrm{IMF}_{\mathrm{p}}$ also exhibits a strong correlation with storm intensity ($\mathrm{R} = 0.723$), it is weaker than the correlation obtained for $\mathrm{B}_{\mathrm{zp}}$ ($\mathrm{R} = -0.811$). This shows that the southward component of the interplanetary magnetic field is more relevant than the total IMF magnitude, consistent with the physical role of southward $\mathrm{B}_{\mathrm{z}}$ in enhancing magnetic reconnection between the interplanetary magnetic field and the Earth's magnetic field.

These findings are consistent with previous studies. In all three studies discussed below, the correlations were calculated using signed $\mathrm{Dst}_{\min}$, rather than $|\mathrm{Dst}_{\min}|$. \citet{gonzalez2005} analyzed 64 storms with $\mathrm{Dst}_{\min} \leq -85$ nT during 1997--2002 and reported strong correlations for $\mathrm{B}_{\mathrm{sp}}$ ($\mathrm{R}=-0.82$), where $\mathrm{B}_{\mathrm{sp}}$ is the positive magnitude of the peak southward IMF and therefore $\mathrm{B}_{\mathrm{sp}}=-\mathrm{B}_{\mathrm{zp}}$. The correlation was also strong for $\mathrm{E}_{\mathrm{yp}}$ ($\mathrm{R}=-0.87$), while the correlation for $\mathrm{E}_{\mathrm{yi}}$, integrated up to the time of $\mathrm{B}_{\mathrm{sp}}$, was weaker ($\mathrm{R}=-0.53$). \citet{balachandran2021} analyzed 170 storms with $\mathrm{Dst}_{\min}<-100$ nT during 1967--2018 and found correlations of $\mathrm{R}=0.73$ ($\mathrm{N}=130$) for $\mathrm{B}_{\mathrm{zp}}$ and $\mathrm{R}=0.56$ ($\mathrm{N}=124$) for the negative integral of $\mathrm{E}_{\mathrm{y}}$ during the main phase. \citet{echer2008} analyzed 90 storms with $\mathrm{Dst}_{\min} \leq -100$ nT during 1996--2006 and also reported strong correlations for $\mathrm{B}_{\mathrm{zp}}$ ($\mathrm{R}=0.80$) and $\mathrm{E}_{\mathrm{yp}}$ ($\mathrm{R}=-0.84$).

For stronger storms ($|\mathrm{Dst}_{\min}| \geq 200$ nT), the present analysis gives a stronger correlation for $\mathrm{E}_{\mathrm{yi}}$ ($\mathrm{R} = 0.642$) than for $\mathrm{E}_{\mathrm{yp}}$ ($\mathrm{R} = 0.584$). This suggests that, for stronger storms, peak $\mathrm{E}_{\mathrm{y}}$ alone may not capture storm intensity as effectively as sustained solar wind driving. This is physically reasonable because stronger storms require a substantial build-up of the ring current during the main phase. A sustained $\mathrm{E}_{\mathrm{y}}$ driver, represented by the time-integrated $\mathrm{E}_{\mathrm{y}}$ parameter, can show a stronger relationship with storm intensity than the peak $\mathrm{E}_{\mathrm{y}}$ value alone. This result is consistent with \citet{watari2023}, who analyzed 18 intense geomagnetic storms during 1996--2021 with $|\mathrm{Dst}_{\min}| > 200$ nT and found a stronger correlation for $\mathrm{E}_{\mathrm{yi}}$ ($\mathrm{R} = 0.838$) than for $\mathrm{E}_{\mathrm{yp}}$ ($\mathrm{R} = 0.586$). A direct application of the present framework to the specific storm sample used by \citet{watari2023} is presented in \ref{app:watari}.

Several limitations remain. The main phase onset is based on an operational estimate and may remain ambiguous for highly complex storms. In addition, missing OMNI values substantially reduce the final storm sample used in the solar wind analysis: from 1643 filtered storms, only 990 remain data-complete. This mainly affects older storms, where missing values are more frequent in the available measurements.

\section{Conclusions}
\label{sec:conclusions}

A web application was developed for retrospective detection and analysis of geomagnetic storms using hourly Dst and OMNIWeb solar wind data. The application detects past storms, identifies their main phases, computes solar wind derived parameters during the main phase, and evaluates their association with storm intensity.

Using the selected default parameters, the application detected 1796 storms from 1964 to May 2026, consisting of 1422 moderate storms, 347 intense storms, and 27 super-storms. After filtering for reliable main phase identification, 1643 storms were retained. In this filtered sample, the average main phase duration per class was 9.25 hours for moderate storms, 11.30 hours for intense storms and 13.75 hours for super-storms. After applying the additional requirement of complete OMNIWeb solar wind data availability, the sample was reduced to 990 storms, which were then used to examine the relationship between storm intensity and solar wind derived parameters over the main phase.

For the data-complete sample of 990 storms with $|\mathrm{Dst}_{\min}|\geq 50$ nT, the strongest correlations with $|\mathrm{Dst}_{\min}|$ were found for $\mathrm{E}_{\mathrm{yp}}$ ($\mathrm{R}=0.836$), $\mathrm{B}_{\mathrm{zp}}$ ($\mathrm{R}=-0.811$), and $\mathrm{E}_{\mathrm{yi}}$ ($\mathrm{R}=0.759$). These results support the importance of the solar wind electric field and southward IMF in storm intensification. In contrast, $\mathrm{V}_{\mathrm{p}}$ and main phase duration exhibited weak correlations with storm intensity ($\mathrm{R}=0.288$ and $\mathrm{R}=0.115$, respectively). However, when the analysis was restricted to storms with $|\mathrm{Dst}_{\min}|\geq 200$ nT, $\mathrm{E}_{\mathrm{yi}}$ exhibited a stronger correlation with storm intensity than $\mathrm{E}_{\mathrm{yp}}$. This suggests that, for stronger storms, sustained solar wind driving over the main phase is more closely associated with storm intensity than peak $\mathrm{E}_{\mathrm{y}}$.

Overall, the application provides a flexible and reproducible framework for geomagnetic storm detection, main phase identification, and solar wind association analysis. By allowing the algorithm parameters to be adjusted and by exporting all results as CSV files, the tool supports further user-defined analysis.

\section{App availability}

The web application is available at: \url{https://geomagnetic-storm-mainphase-analysis.streamlit.app/}

GitHub repository: \url{https://github.com/Daniil-Papakonstantinou/magnetic_storm_analysis-app}

A frozen archive of the application version (v1.0.0) used in this paper is available on Zenodo: \url{https://doi.org/10.5281/zenodo.22281855}
\vfill
\noindent\textit{An earlier version of this work was presented as a poster at the 16th Quadrennial Solar-Terrestrial Physics Symposium, 2026.}

\clearpage
\appendix
\setcounter{table}{0}
\renewcommand{\thetable}{A\arabic{table}}
\renewcommand{\thesection}{\Alph{section}}
\makeatletter
\renewcommand{\@seccntformat}[1]{%
  \ifnum\pdfstrcmp{#1}{section}=0 Appendix~\csname the#1\endcsname\quad
  \else\csname the#1\endcsname\quad\fi}
\makeatother

\section{Reproduction of the Watari et al. storm sample}
\makeatletter
\protected@edef\@currentlabel{Appendix~\thesection}
\makeatother
\label{app:watari}

\citet{watari2023} analyzed 18 geomagnetic storms with $|\mathrm{Dst}_{\min}| > 200$ nT during 1996--2021. To reproduce their event sample, the application was run from 1 January 1996 to 1 January 2022 with STORM\_LEVEL set to $-201$ nT. Stage 1 detected the same 18 storms, with minimum Dst values and timestamps matching those reported in their study, as shown in Table~\ref{tab:watari-storms}.

\begin{table}[htbp]
\centering
\setlength{\tabcolsep}{18pt}
\caption{All 18 storms with $|\mathrm{Dst}_{\min}| > 200$ nT during 1996--2021 detected by the application, matching the storm sample used by \citet{watari2023}.}
\label{tab:watari-storms}
\begin{tabular}{lcr}
\toprule
\textbf{No} & {\bfseries\boldmath $\mathrm{t}_{\min}$ (UT)} & {\bfseries\boldmath $\mathrm{Dst}_{\min}$ (nT)} \\
\midrule
1 & 1998-05-04 05:00 & $-205$ \\
2 & 1998-09-25 09:00 & $-207$ \\
3 & 1999-10-22 06:00 & $-237$ \\
4 & 2000-04-07 00:00 & $-292$ \\
5 & 2000-07-16 00:00 & $-300$ \\
6 & 2000-08-12 09:00 & $-234$ \\
7 & 2000-09-17 23:00 & $-201$ \\
8 & 2001-03-31 08:00 & $-387$ \\
9 & 2001-04-11 23:00 & $-271$ \\
10 & 2001-11-06 06:00 & $-292$ \\
11 & 2001-11-24 16:00 & $-221$ \\
12 & 2003-10-30 00:00 & $-353$ \\
13 & 2003-10-30 22:00 & $-383$ \\
14 & 2003-11-20 20:00 & $-422$ \\
15 & 2004-11-08 06:00 & $-374$ \\
16 & 2004-11-10 10:00 & $-263$ \\
17 & 2005-05-15 08:00 & $-247$ \\
18 & 2015-03-17 22:00 & $-234$ \\
\bottomrule
\end{tabular}
\end{table}
\FloatBarrier

Stage 2 excluded storms 13 and 16 due to previous-storm overlap, leaving 16 storms in the filtered sample. Storms 10 and 12 contained multiple missing solar wind-speed values and therefore also had incomplete $\mathrm{E}_{\mathrm{y}}$ data, reducing the Stage 3 data-complete sample to 14 storms (Storms 12 and 13 are the storms shown in \hyperref[ex:example1]{Example 1}). The $\mathrm{B}_{\mathrm{zp}}$ correlation was calculated using all 16 filtered storms because they were data-complete in terms of $\mathrm{B}_{\mathrm{z}}$, whereas the $\mathrm{E}_{\mathrm{yp}}$ and $\mathrm{E}_{\mathrm{yi}}$ correlations used the 14-storm Stage 3 sample. The results are shown in Table~\ref{tab:application-watari-correlations}.

\begin{table}[htbp]
\centering
\setlength{\tabcolsep}{18pt}
\caption{Correlations of $\mathrm{B}_{\mathrm{zp}}$, $\mathrm{E}_{\mathrm{yp}}$, and $\mathrm{E}_{\mathrm{yi}}$ with $|\mathrm{Dst}_{\min}|$ calculated by the application}
\label{tab:application-watari-correlations}
\begin{tabular}{lcc}
\toprule
\textbf{Parameter} & {\bfseries\boldmath $\mathrm{R}$} & {\bfseries\boldmath $\mathrm{N}$} \\
\midrule
$\mathrm{B}_{\mathrm{zp}}$ & $-0.661$ & 16 \\
$\mathrm{E}_{\mathrm{yp}}$ & $0.542$ & 14 \\
$\mathrm{E}_{\mathrm{yi}}$ & $0.785$ & 14 \\
\bottomrule
\end{tabular}
\end{table}
\FloatBarrier

To fill gaps in the OMNI plasma data, \citet{watari2023} calculated hourly solar wind-speed values using alpha-particle speeds measured by ACE/SWICS and then calculated the corresponding $\mathrm{E}_{\mathrm{y}}$ values using those speeds and the OMNI magnetic field data. The correlations reported in that study are shown in Table~\ref{tab:watari-correlations}. For consistency with the present application, their $\mathrm{B}_{\mathrm{sp}}$ correlation of $\mathrm{R}=0.579$ with $|\mathrm{Dst}_{\min}|$ was expressed using the signed $\mathrm{B}_{\mathrm{zp}}$ convention. Since $\mathrm{B}_{\mathrm{sp}}$ is the positive magnitude of the peak southward IMF, $\mathrm{B}_{\mathrm{zp}}=-\mathrm{B}_{\mathrm{sp}}$.

\begin{table}[htbp]
\centering
\setlength{\tabcolsep}{18pt}
\caption{Correlations of $\mathrm{B}_{\mathrm{zp}}$, $\mathrm{E}_{\mathrm{yp}}$, and $\mathrm{E}_{\mathrm{yi}}$ with $|\mathrm{Dst}_{\min}|$ reported by \citet{watari2023}.}
\label{tab:watari-correlations}
\begin{tabular}{lcc}
\toprule
\textbf{Parameter} & {\bfseries\boldmath $\mathrm{R}$} & {\bfseries\boldmath $\mathrm{N}$} \\
\midrule
$\mathrm{B}_{\mathrm{zp}}$ & $-0.579$ & 18 \\
$\mathrm{E}_{\mathrm{yp}}$ & $0.586$ & 18 \\
$\mathrm{E}_{\mathrm{yi}}$ & $0.838$ & 18 \\
\bottomrule
\end{tabular}
\end{table}
\FloatBarrier

In both analyses of this storm sample, the time-integrated \(\mathrm{E}_{\mathrm{y}}\) over the main phase showed the strongest correlation with \(|\mathrm{Dst}_{\min}|\). This indicates that sustained solar wind driving is particularly strongly associated with the intensity of stronger geomagnetic storms.

\clearpage
\phantomsection
\pdfbookmark[1]{References}{sec:references}

\end{document}